\documentclass[conference]{IEEEtran}
\usepackage{cite}
\usepackage{amsmath,amssymb,amsfonts}
\usepackage{algorithmic}
\usepackage{graphicx}
\usepackage{textcomp}
\usepackage{amsmath}
\usepackage{xcolor}

\usepackage{fancyhdr} % Load the package

\def\BibTeX{{\rm B\kern-.05em{\sc i\kern-.025em b}\kern-.08em
    T\kern-.1667em\lower.7ex\hbox{E}\kern-.125emX}}
\begin{document}

\title{Developments Towards Superconducting On-Chip Fourier Transform Spectrometers (SOFTS)}

\author{N. Klimovich$^{\text{1},*}$, R. Basu Thakur$^{\text{2,3} ,\star}$, Dan Cunnane$^\text{2}$, Paul Goldsmith$^\text{2}$, Phil Mauskopf$^\text{4}$, Peter K. Day$^\text{2}$,  B-.K. Tan$^\text{1}$\\
$^\text{1}$\small\textit{Department of Physics (Astrophysics), University of Oxford, United Kingdom}\\
$^\text{2}$\textit{Jet Propulsion Laboratory, Pasadena, United States of America}\\
$^\text{3}$\textit{California Institute of Technology, Pasadena, United States of America}\\
$^\text{4}$\textit{Arizona State University, Phoenix, United States of America}}

\maketitle
\thispagestyle{fancy}
\begin{abstract}
\normalsize
We need spectroscopic imaging at sub-millimetre wavelengths to answer fundamental questions about the origins of our universe, structure formation and galaxy evolution. Often, there are steep trade-offs between the number of detectors and their spectral resolving power in completing such 3D surveys. Novel astrophotonics technologies, such as the Superconducting On-Chip Fourier Transform Spectrometer (SOFTS), are underway to break instrumentation barriers and enable the packing of a focal-plane array with many high spectral resolution submillimeter detectors. Here, we summarize recent advances in SOFTS development, focusing on potential spectroscopic resolution and packing density within a focal plane array. \let\thefootnote\relax
\footnotetext{* nikita.klimovich@physics.ox.ac.uk, $\star$ ritoban.basu.thakur@jpl.nasa.gov}
\end{abstract}


\section{Introduction}

%Many of the 
There are many gaps in our understanding of the formation of galaxies in the early universe and their subsequent evolution through cosmic time. The data necessary to answer these questions lie in the far-infrared spectroscopic signals which tell us the redshift (and thus distance) of sources, the kinematics of individual galaxies, and their chemical composition. For the most pressing science cases, simultaneous spatial and spectral resolution is further required to construct spatial--velocity maps and remove foreground signals, which dramatically increase survey time, as present-day high-resolution spectrographs can generally measure the spectra of only one point in the sky at a time. 

Historically, the physical scale of millimeter (mm) and sub-mm wave spectrometers has limited the possibility of packing them into a focal plane array. Although SIS receiver development is moving towards that direction, SIS mixers can only simultaneously cover a small portion of the RF range and thereby needs time-consuming local oscillator tuning to scan over the frequencies required in many astronomical observations.\cite{Tan_2025} For the many applications that do not need as high of spectral resolution as offered by SIS mixers but require simultaneous coverage of a much wider RF range covering several 100s GHz, other technological solutions are required. 

Recent advances in astrophotonics technology enable instruments to push past this limitation, with emerging miniaturized filterbank, quasi-grating, and Fourier transform spectrometers \cite{astrophotonicsRoadmap}. The Superconducting On-Chip Fourier Transform Spectrometer (SOFTS) \cite{SOFTSinitial} is notable among these emerging technologies due to its uniquely field-tunable spectral resolution and especially small spectro-imaging pixel (spaxel) size. Since the mapping speed for large surveys scales with the number of detectors \cite{astrophotonicsRoadmap}, SOFTS can outperform filterbank spectrometers for large field of view telescopes due to the higher number of spaxels that can be packed within the focal plane. 

SOFTS is formed with two superconducting transmission lines (STLs) sandwiched between two hybrid couplers. By utilizing the current-dependent propagation velocity of the high kinetic inductance STLs, similar to the operation of travelling wave parametric amplifiers (TWPAs),\cite{Tan_2024} SOFTS modulates phase delay within each arm of its interferometer without the need for mirrors, wire-grid beam-splitters etc. Hence, its operation is entirely electronically and does not require any optomechanical components. The inherently low phase velocities of the high kinetic inductance STLs (with $v < 0.01c$), result in a corresponding shrinking of the spectrometer size, with a spectral resolution of $R \sim 10^3$ achievable within a much smaller physical footprint than the area of a spaxel's horn aperture. 

\begin{figure}[h]
    \centering
    \includegraphics[width=0.48\textwidth]{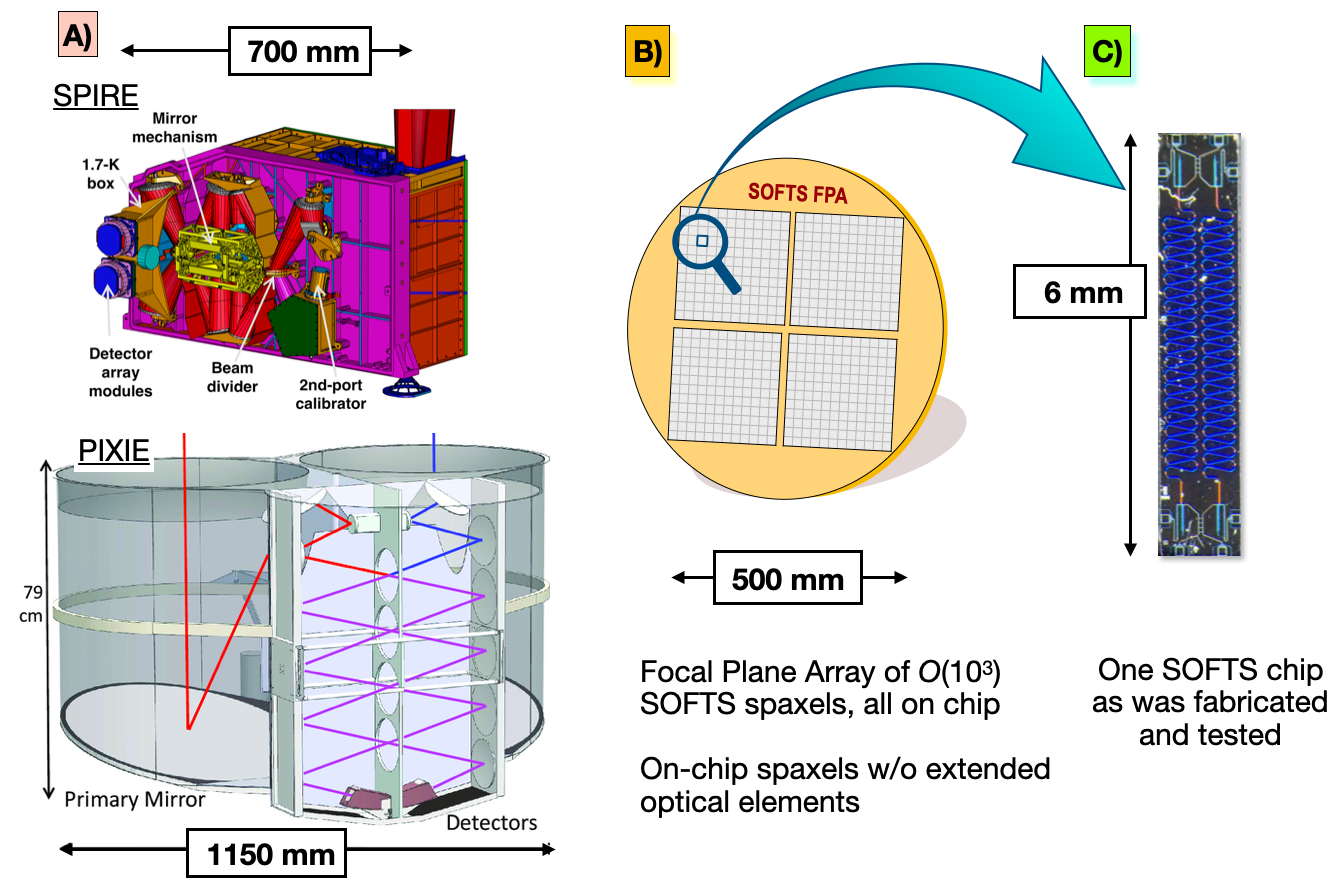}
    \caption{\textbf{A)} State-of-the-art FTSs are generally meter-scale instruments (e.g., SPIRE, PIXIE, adapted from ~\cite{herschel_fts, PIXIE}), even with only a few spectro-imaging pixels (\emph{spaxels}). \textbf{B)} Thousands of SOFTS' \emph{spaxels} can be arrayed to realize revolutionary focal plane arrays less than meter-scale size, covering GHz--THz frequency range with spectral resolving power $\mathbf{R} \gtrsim 10^3$. \textbf{C)} An all-electronic SOFTS that was built and tested at Ka-band frequencies. It is $100\times$ smaller than the opto-mechanical counterparts.}
    \label{fig:grandopening}
\end{figure}

This paper describes the projected performance, scale, and observational capabilities of a SOFTS focal plane array, based on laboratory measurements of our prototype devices designed and tested in the 30-40 GHz frequency range. 

\section{Ka-band (26--42\,GHz) SOFTS}

\begin{figure}[h]
    \centering
    \includegraphics[width=0.48\textwidth]{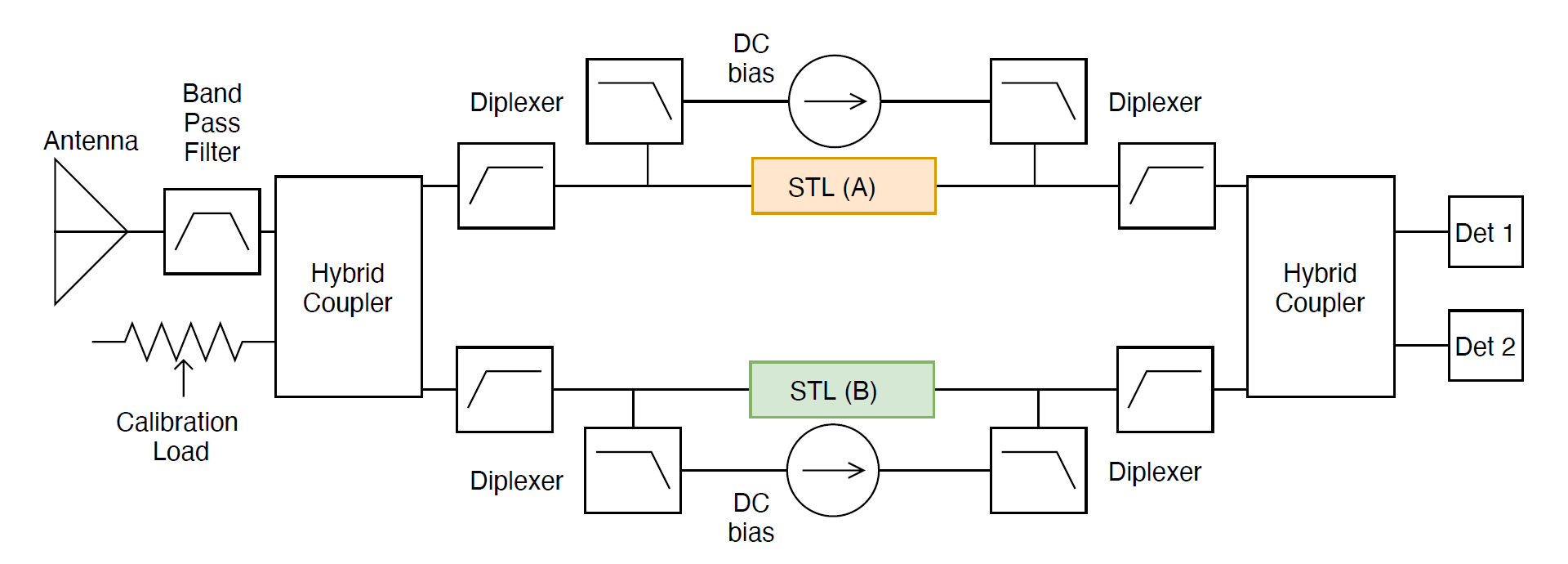}
    \caption{A schematic diagram of a single SOFTS pixel comprising a broadband antenna, two hybrid couplers and two high kinetic inductance superconducting transmission lines (STLs) biased with DC current to alter the phase delay of the STLs. There is great flexibility in the choice of antennae at the input and detectors at the output.}
    \label{fig:SOFTS_diagram}
\end{figure}

\begin{figure}[h]
    \centering
    \includegraphics[width=0.48\textwidth]{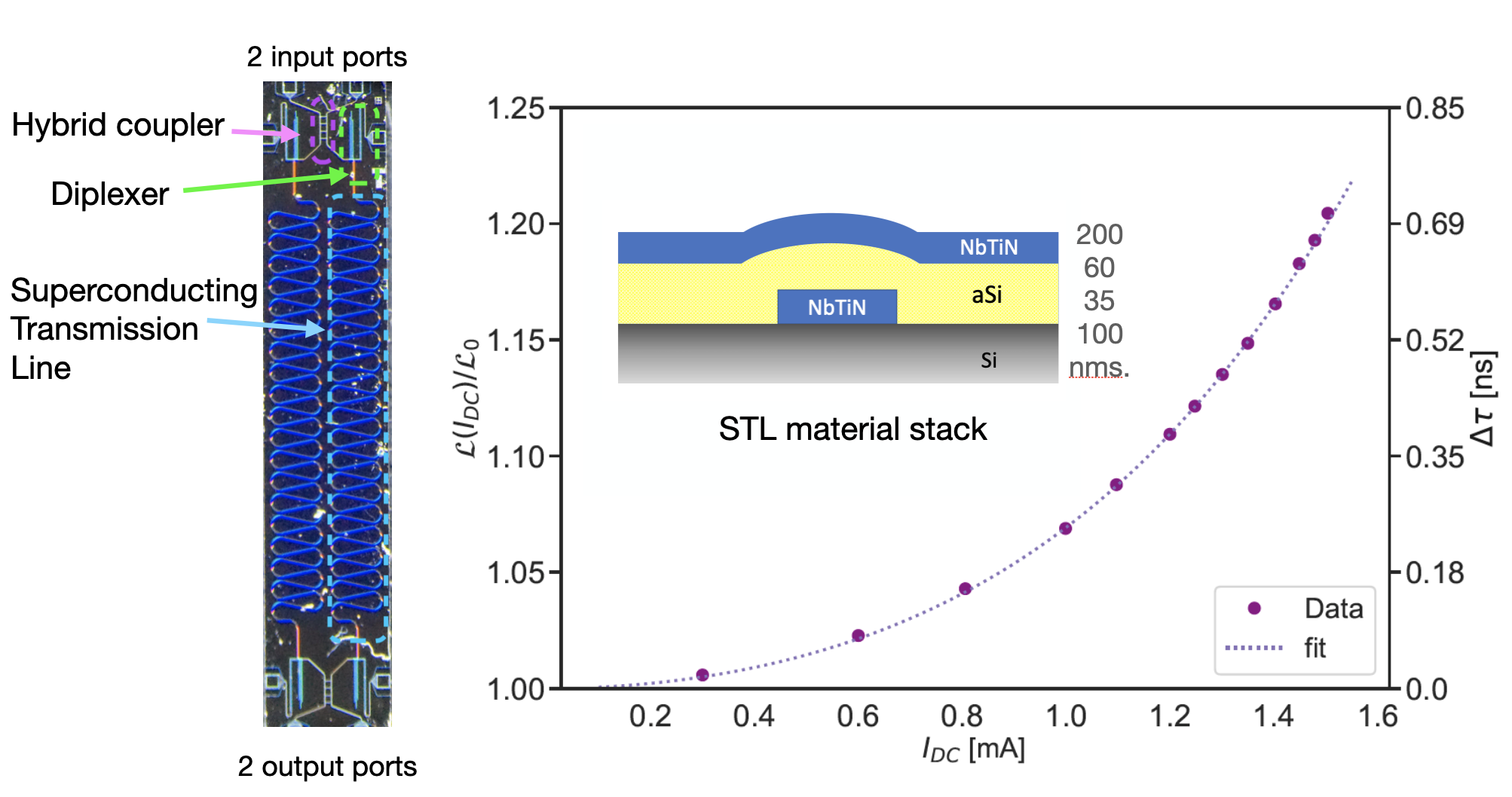}
    \caption{\textbf{Left:} Photograph of a Ka-band SOFTS' pixel chip consisting of hybrid couplers, diplexers, and superconducting transmission lines (STLs). \textbf{Right:} Fabrication stack of the STL with the corresponding materials and thickness presented, and the experimental result of a single STL showing 0.7\,ns maximum delay with an applied DC current of 1.5\,mA.}
    \label{fig:ka_image}
\end{figure}

Fig. \ref{fig:SOFTS_diagram}, adapted from \cite{SOFTS}, shows the initial 1--10\,GHz SOFTS contructed using modular commercial microwave components alongside the nonlinear kinetic inductance superconducting transmission lines made out of niobium titanium nitride (NbTiN) to construct the interferometric architecture \cite{SOFTSinitial}, which is similar to the balanced TWPA scheme proposed for reading out SIS mixers \cite{JoeBalanced}. We presented the preliminary experimental verification results in \cite{SOFTSinitial}. While those preliminary results experimentally verified the feasibility of the SOFTS' architecture, on-chip integration of all required microwave components is necessary to realise an array that can densely fills a focal plane. To that end, we designed and fabricated a Ka-band SOFTS as an integrated on-chip 4-port Mach-Zehnder interferometer. On-chip components such as the hybrid couplers for beam splitting/combining, the diplexers for combining RF signal and DC bias current to the STLs were designed and simulated~\cite{SOFTS} using commercial Sonnet and Ansys HFSS EM simulation softwares. The fabricated chip is shown in Fig.\,\ref{fig:ka_image} with a 6\,mm$\times$1\,mm size footprint primarily to encompass the two 21\,mm long (when unwrapped) STL lines. The right panel of the figure shows the measured phase delay response of the STL corresponding to the strength of the DC biasing current.

\begin{figure}[h]
    \centering
    \includegraphics[width=0.48\textwidth]{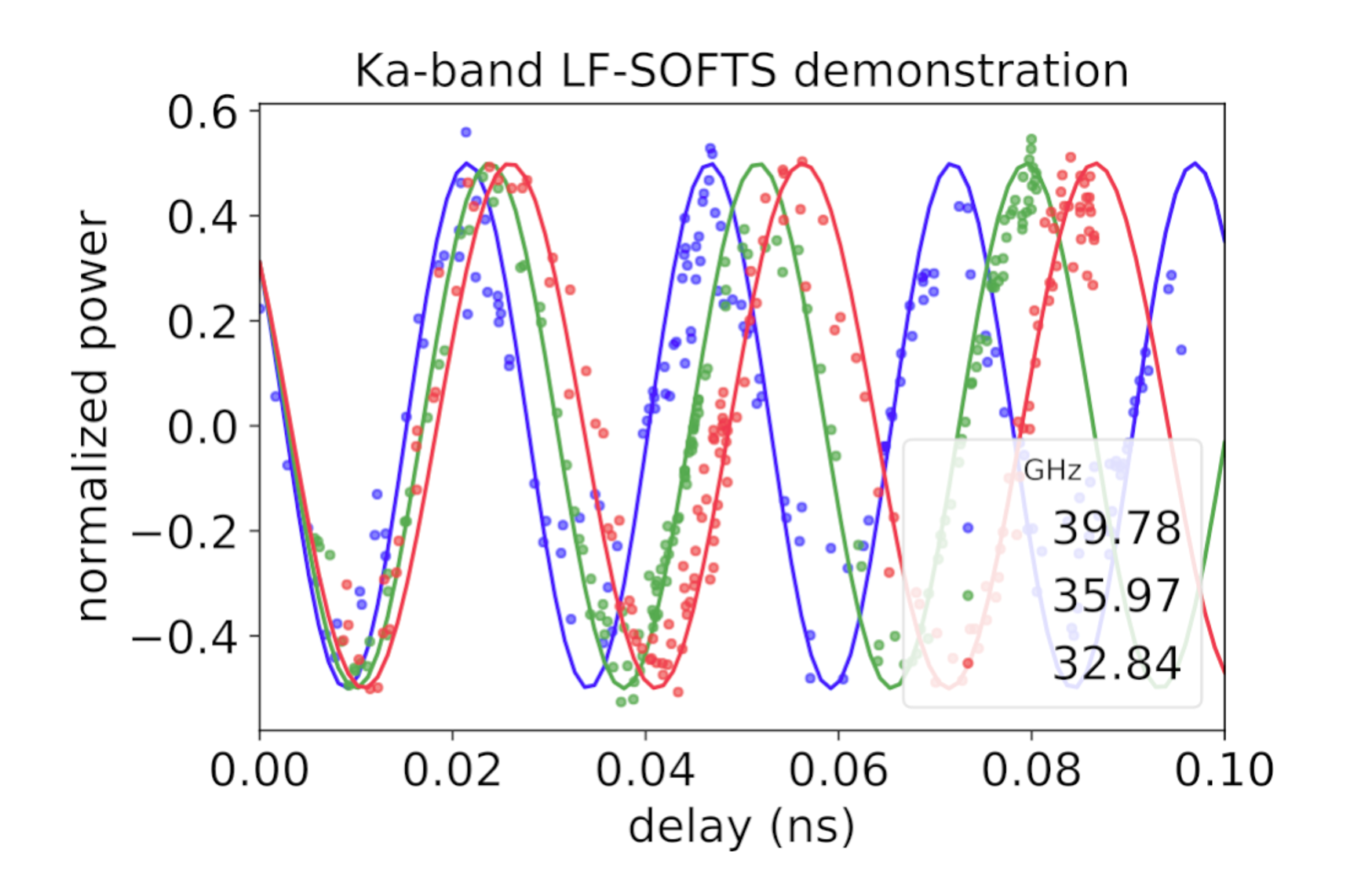}
    \caption{Ka-band SOFTS' demonstration with three single-tone interferograms zoomed in on the first 100\,ps of delay. Circuit-simulated theoretical fits are shown in solid-lines, having the frequencies agree within $\sim$1\% error of the experimentally supplied tones. Amplitude noise in the experimental data is resulted from the commercial interconnect components used in laboratory testing setup and are not originated from the SOFT device.}
    \label{fig:ka_exp}
\end{figure}

Sub-kelvin tests of the Ka-band SOFTS enabled characterizing their performance as interferometers. Fig.\,\ref{fig:ka_exp} shows that for single tone input, we successfully obtain cosine interferograms when adjusting the delay time in one of the arm by applying a DC bias current. The ability to easily distinguish and accurately recover the frequencies of each single-tone input at various frequencies is the basis for broadband spectroscopic performance characterization and clearly demonstrates the SOFTS' operational principle experimentally \cite{SOFTS,steiger_thesis}. However, film's smoothness variations and impedance mismatches between on-chip components led to somewhat limited performances, compared to what was expected based on the performance of individual Ka-band transmission lines shown in Fig. \ref{fig:ka_image}, adapted from \cite{Shu}. These results, with a first-of-its-kind prototype, are nonetheless a major step towards realizing a single usable SOFTS spaxel.

% Unfortunately, poor quality of the NbTiN film within this particular fabrication run greatly diminished the achievable resolution of the on-chip integrated SOFTS chips compared to what we would expect based on the earlier performance of the individual reference Ka-band transmission lines shown in Fig. \ref{fig:ka_image}. These results are nonetheless a major step towards realizing a single usable SOFTS spaxel.--> altered, RBT.

\section{Spectral Resolution \& Upper Frequency Limit}

The frequency resolution of an FTS is inversely proportional to the maximum time delay difference between the two arms of the interferometer of length $L$ and is given by 
% I changed the variable name to match eq 3. -> RBT
\begin{equation}
\delta \nu = \frac{2v_\text{ph}}{L} \sqrt{\frac{\mathcal{L}(0)}{\mathcal{L}(I_c) - \mathcal{L}(0)}},
\end{equation}
\noindent where $v_{ph}$ is the phase velocity of the STL, and $\mathcal{L}(I)$ is the nonlinear inductance as a function of current. This kinetic inductance change due to the applied DC current results in a different effective electrical length for the STL, which is here expressed as a ratio that scales with the physical length $L(0)$ and thereby sets the maximum $\Delta L$ in the standard FTS resolution relation. This dependence is typically expressed as a series expansion: 
\begin{equation}
\mathcal{L}(I) = \mathcal{L}_0 \left[1 + \left(\frac{I}{I_*}\right)^2  + \left(\frac{I}{I_*'}\right)^4 + \ ... \right]
\end{equation}
\noindent where $I_c$ is the critical current of the superconducting film, and $I_*, I_*'$ are the nonlinearity scale parameters. The ratios $I_c/I_*$ and $I_c/I_*'$ are typically reduced by defects within the superconducting film which locally reduce the critical current.

Typical values of $I_c/I_*$ for a long devices ($L > 100$mm) are measured to be around 0.2; however, values as high as 0.35 have been measured in shorter devices ($L \simeq 21$mm) with fewer defects \cite{Shu}. This ratio of critical current directly corresponds to the maximum percentage we can shift the kinetic inductance (and consequently, the phase delay) for a certain DC driving current value. Combined with the length of the Ka-band SOFTS' STL, these values have resulted in a presently experimentally obtainable SOFTS resolution of $0.26$\,GHz at Ka-band. Replacing the NbTiN film with a higher kinetic inductance material such as magnesium diborite (MgB$_2$) film, and further effort in increasing the winding factor of the STL (see \cite{Farzad}), we expect it would allow for the construction of SOFTS with $100$ MHz resolution within a 5\,mm$\times$5\,mm physical footprint. 

The maximum detectable frequency of a SOFTS spaxel is set by the gap of the superconducting material used. For NbTiN, this results in a potential operation up to approximately 1.5\,THz. This could be significantly increased to near 3\,THz using MgB$_2$, but more study is needed to determine the frequency dependence of Cooper pair breaking against de-pairing in such dual-gap materials. MgB$_2$ films offer further advantages due to its higher kinetic inductance of approximately 28 pH/$\scriptscriptstyle{\square}$ compared to the 7 pH/$\scriptscriptstyle{\square}$ in our NbTiN films, while maintaining similar values of $I_c/I_*$ \cite{mgb2,changsub}. As a result, MgB$_2$-based SOFTS can theoretically provide similar resolution with 4$\times$ shorter devices than an equivalent NbTiN counterpart. This reduction of device length has a compounding benefit as it reduces the likelihood of defects that lower $I_c/I_*$ and therefore provide an even greater benefit in practice. Nevertheless, further research and development will be required to fully capture the capabilities and limits of such THz-enabled SOFTS technology. 

\section{Detector Size}

\begin{figure}[h]
\centering
\includegraphics[height=5cm]{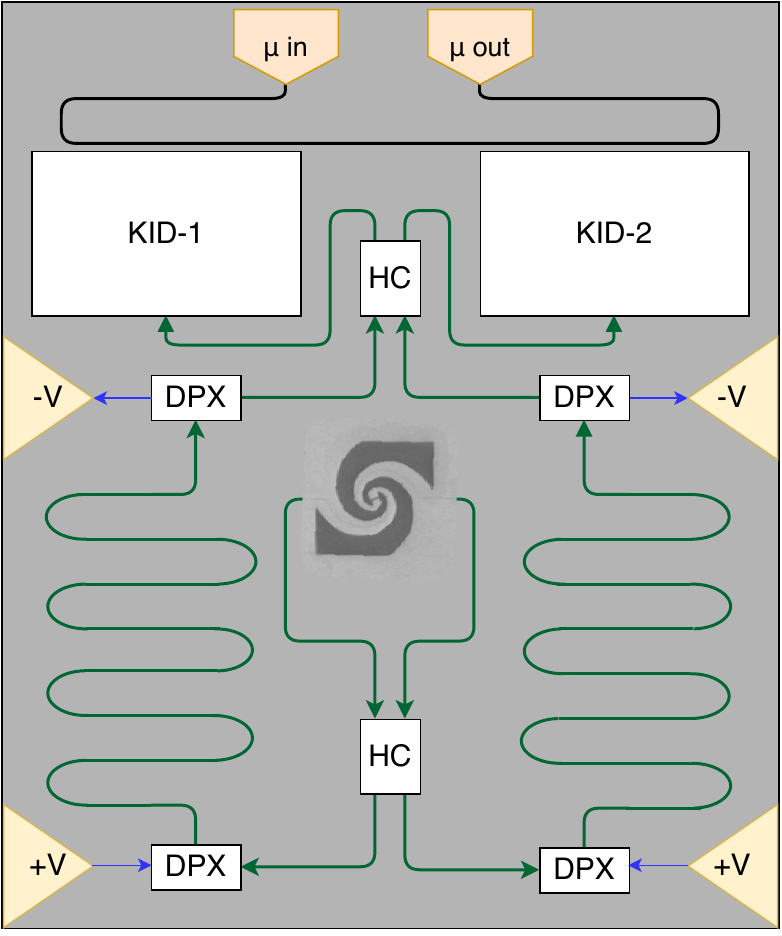}%{SOFTSspaxel.png}
\caption{A schematic of a realizable SOFTS spaxel consisting of a broadband spiral antenna, two hybrid coupler, four diplexers, and the superconducting transmission lines (STL) arms, with the kinetic inductance detectors (KIDs) readout. KIDs readout feedline with bond-pads are marked as $\mu$-wave (only $\mu$ without wave in the fiture) in /out lables and and the STL biasing current is supplied via the $\pm$V bond pads.
}
\label{fig:softs_spaxel}
\end{figure}

\begin{figure}[h]
\centering
\label{fig:softs_area}
\includegraphics[width=8cm]{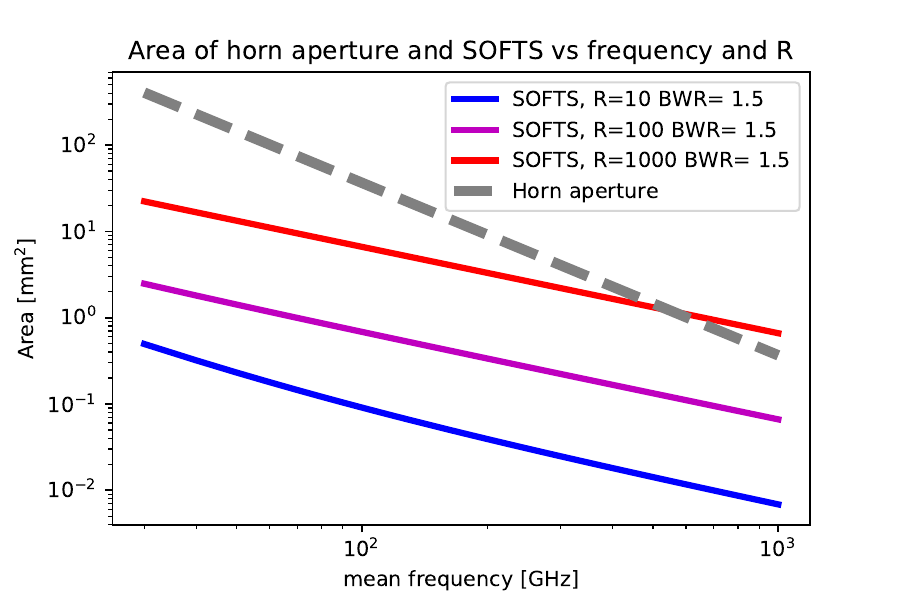}
\caption{Area occupied by the required horn aperture compared to the NbTiN SOFTS devices with varying $R$. Modelling is done for a bandwidth ratio of 1.5. For smaller $R$, a shorter STL length would be required, hence reducing the overall dimensions. Detector area is assumed to be $50\mu$m $\times 200\mu$m.} 
\end{figure}

In a focal plane array consisting of SOFTS spaxels, the packing density is generally not compromised when targeting resolutions of $\mathbf{R}\sim10^3$ (higher $\mathbf{R} \Rightarrow$ longer STLs). A potential layout for a single spaxel shown in Fig.~\ref{fig:softs_spaxel} uses meandered STLs to realize a small footprint (6 mm $\times$5 mm) device operating at 600 GHz frequencies. We compare this area in the focal plane array to the horn aperture required for each detector in Fig.~\ref{fig:softs_area}. Even for an NbTiN SOFTS, the packing density of the spaxels in the focal plane will be limited mainly by the size of the horn apertures at frequencies below 600\,GHz; with this limit improving at all operating frequencies for MgB$_2$-based SOFTS.

\section{Measurement Capabilities}

\subsection{Observational flexibility \& sensitivity}
When unbiased, SOFTS acts as a standard single-mode antenna-coupled bolometer since identical STL lengths will couple all input power onto one of the detectors. Thus, it is capable of performing a photometric survey with practically instantaneous spectroscopic follow-up measurements. This enables a great deal of observational flexibility for large surveys as each spaxel in the focal plane array can be individually tuned between a bolometric and spectroscopic mode depending on the target that the spaxel is observing. 

The broadband capabilities of the FTS architecture introduce an inherent tradeoff as the photon noise (noise equivalent power, NEP) across the entire observable bandwidth will be detectable by the bolometric detectors, e.g., the MKIDs detector shown in Fig.\,\ref{fig:softs_spaxel}. On the other hand, the effect of the noise on the reconstructed spectrum is given by the noise equivalent spectral brightness (NESB) and is notably not dependent on the number of spectral channels \cite{PIXIE} i.e.,
\begin{equation}
\text{NESB}_\text{array} = \frac{2\text{NEP}_\text{photon}}{A\Omega(\nu) \cdot \delta \nu \cdot \sqrt{T_\text{obs} N_\text{spaxels}}}.
    \label{eq:nesb_array}
\end{equation}

For each SOFTS' spaxel, this means only a pair of KID detectors with a typical NEP$_\text{det}\approx 10^{-18}/\sqrt{\text{Hz}}$ is needed. This is another clear advantage of SOFTS architecture over, for e.g., filter-bank spectrometers which require fabrication of thousands of ultra sensitive state-of-the-art bolometric detectors with NEP$_\text{det} < 10^{-19}/\sqrt{\text{Hz}}$, efficiently  coupled to each spectral channel, within a single spaxel. 

This relatively high per-pixel NEP compared to filterbank technologies is commonly used to argue against the efficacy of FTS for astronomical surveys; however, several other factors need to be considered to determine mapping speed. Most notably, the high degree of spatial multiplexing enabled by packing a focal plane array counteracts this traditional disadvantage. Fig. 18 in chapter 10 of ~\cite{astrophotonicsRoadmap} compares the mapping speed of SOFTS with respect to filterbanks and other on-chip spectrometer technologies to conclude that for spectro-imaging of extended objects, enabled by SOFTS, is a massive advantage that is unparalleled by other technologies. Specifically, mapping speed is $\propto$ the total number of detectors. Since SOFTS uses only 2 detectors per spaxel without size constraints, many more spaxels can be fielded. Thus without sacrificing mapping speed, one can attain broadband spectra over a large field of view, simultaneously with thousands of imaging pixels.

\subsection{Additional enhanced operations}

The small physical footprint and ease of electronic-control of delay time in SOFTS interferometer also allows for several additional enhanced operation modes compared to a traditional optomechanical FTS. 

\begin{itemize}
    \item SOFTS hybridized with filer banks: Since the physical footprint of a SOFTS spaxel is so small, it is possible to place several low-$\mathbf{R}$ filter bank coupled detectors at the output ports of the high-$\mathbf{R}$ SOFTS. This reduces the NEP$_\text{photon}$ seen by each bolometric detector by dividing it among the channels of the filter bank, thereby improving the NESB$_\text{array}$ by an equivalent amount. Thus SOFTS can work \emph{with} co-fabricated filter banks to flexibly realize focal plane arrays with desired mapping speeds.
    
    \item Null out bright lines: Since phase delay in SOFT is controlled entirely electronic, creative opportunities exist with digital signal processing as the mm/sub-mm spectrum (per spaxel) can be modulated into the audio-band with current scanning. One such scheme can be used to null out bright emission lines. For example, if an object under observation has a very bright emission line at $\nu_\text{line}$, which is not of observational interest, we can engineer the phase delay to $ 1/(2\nu_\text{line})$ that would lead to a destructive interference at $\nu_\text{line}$ ($\cos(2\pi \nu_\text{line}\tau)=-1$). Now, as we vary $\tau$ around this set-point, photons from this line emission will therefore not be dominating the measurements.  
    
    \item Constructively interfere spectral lines: Using the same reasoning as the previous point, we can take the opposite approach where we rapidly modulate the phase delay not in continuous steps to jump between discrete points, where the desired $\nu_\text{line}$ experiences constructive interference at $\cos(2\pi \nu_\text{line}\tau)=1$. On the other hand, RF signals at other frequencies $\nu_s \neq \nu_\text{line}$ will vary between different degrees of constructive and destructive interference as the bias current jumps between the different set values. If this current modulation is performed faster than the readout speed of the detector, such biasing scheme would act as a bandpass filter around $\nu_\text{line}$ and therefore having the benefit of lowering the effective NEP for such measurements.  
    
    \item Multiplexing: The current-scanning speed can be as fast as $\sim$10 MHz limited by the inductance-capacitance network of the STLs. In contrast, the typical readout speed of the detectors are $\ll$ MHz.  Lab demonstrations of current sweeps have been done at $\sim$10 kHz. Thus the DC biasing that has been discussed so far is not strictly necessary and AC biasing can also be pursued with on-chip diplexers. This opens up the ability to multiplex SOFTS' spaxel biasing using  techniques already developed by the transition-edge sensors (TES) and superconducting quantum interference device (SQUID) communities. 
    
\end{itemize}

\subsection{Example telescope observations}

\begin{figure}[h]
\centering
\includegraphics[width=0.5\textwidth]{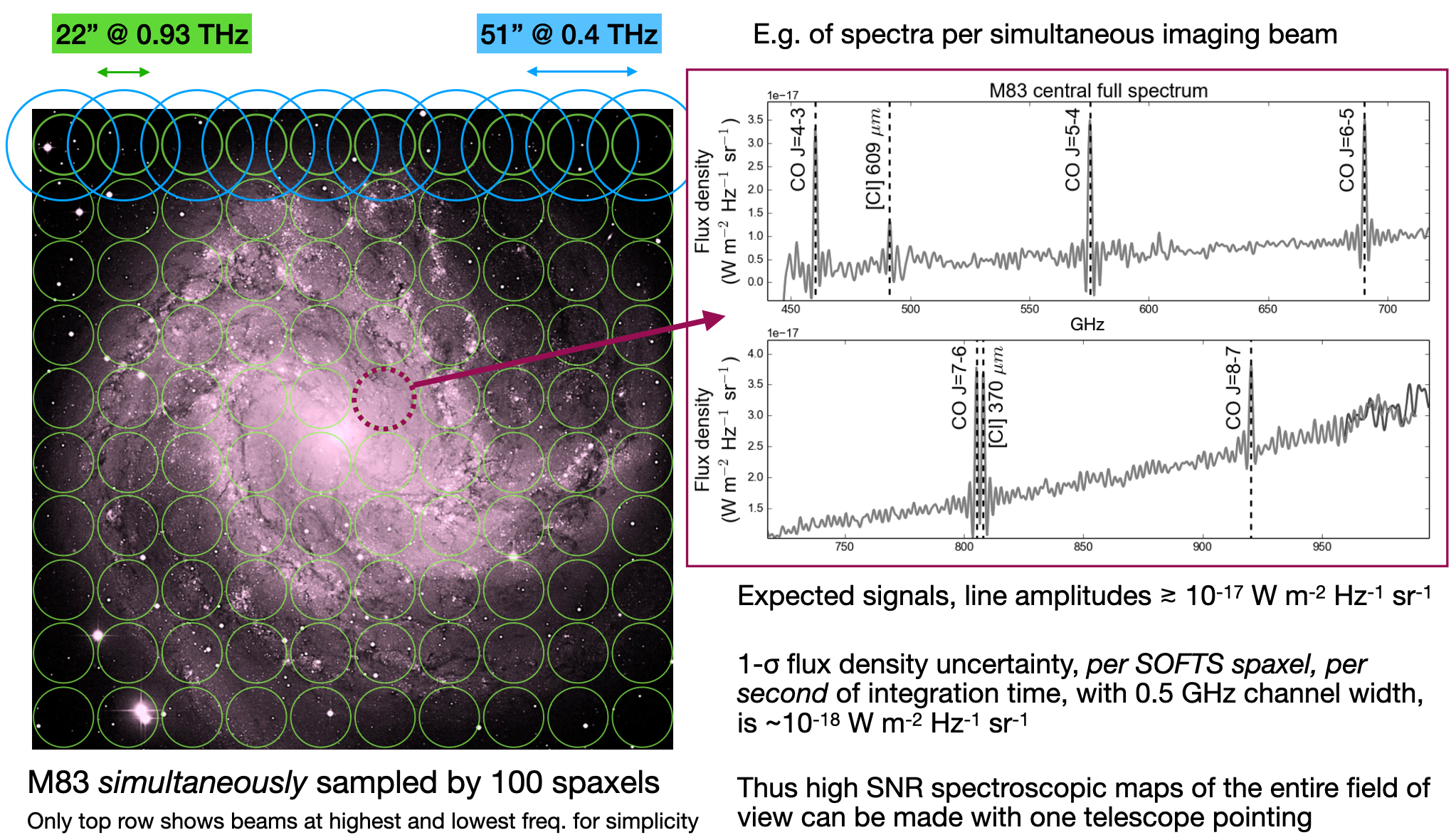}
\caption{Cartoon representation of the M83 imaged with a simulated 10 by 10 SOFTS focal plane array. Each beam empowered by a SOFTS' spaxel captures an individual spectrum with the projected examples shown in the inset (adapted from FTS measurements with Herschel see \cite{wum83}). Sensitivity and image resolution are approximated from ASTRHOS and PRIMA (PRobe far-Infrared Mission for Astrophysics) ~\cite{ASTHROS, PRIMA}, but would ultimately depend on telescope design.}
\label{fig:LIM}
\end{figure}

A galaxy such as M83 shows emission over a $>3$'-sized region, and thus a 10$\times$10 array of pixels could be employed to obtain a complete multi-transition image of the Carbon Monoxide (CO) emission, ready for comparison with other star formation tracers, and thus provide a clear picture of the rate of star formation and the reservoir of material for future star formation in the galaxy. As a proof of concept, Fig.\,\ref{fig:LIM} shows two example beams from a 10$\times$10 SOFTS' spaxel focal plane array observing the M83 at 0.93\,THz and 0.4\,THz. With a 2.5\,m diameter telescope (such as that built for ASTHROS, the Astrophysics Stratospheric Telescope for High Spectral Resolution Observations at Submillimeter-wavelengths, a NASA Astrophysics balloon mission \cite{ASTHROS}), at the high frequency end of SOFTS coverage, the beam-size will be at about $\sim$22” FWHM (full-width half-maximum).

The intensity of the CO line near emission peak of M83 is $\approx 3\times 10^{-17}\; \text{W/m}^2\text{/Hz/sr}$, \cite{wum83}. Accounting for photon noise from the cosmic microwave background (CMB), Cosmic Infrared Background (CIB), cold optics and detectors\footnote{Assuming NEP$_\text{det}\approx 10^{-18}/\sqrt{\text{Hz}}$ over the full band.}, for a single SOFTS spaxel operating between 0.4--0.93\,THz, with 40\% end-to-end efficiency, with just 1 second of integration time, the signal to noise ratio is expected to be $>10$. Even with other systematic inefficiencies, it is clear that a significant measurement can be made with a short integration time covering the entire 0.4--0.93\,THz frequency range which cover an impressive 5 rotational transitions of CO simultaneously with a resolution of 0.5\,GHz. This equivalent to a resolving power, R $\simeq$ 1000, or equivalently a velocity resolution $\delta v$ $\simeq$ 300 kms$^{-1}$. This resolution is sufficient to separate each transition of the $^{13}$CO isotopologue from its corresponding $^{12}$CO line (separation $\simeq$1 part in 24).

\section{Conclusion}

We present the recent advancements in the development of Superconducting On-Chip Fourier Transform Spectrometers (SOFTS). Initial laboratory testing at sub-40 GHz frequencies demonstrate the efficacy of using superconducting nonlinear kinetic inductance to modulate the effective electrical length of the spectrometer arms without the need for optomechanical components. The small scale of these devices allows for the construction of high-resolution ($R > 1000$) spaxels with mm-scale physical footprints and thereby enable future compact focal plane arrays at mm and sub-mm frequencies.

\section*{Acknowledgements}
We thank the Jet Propulsion Laboratory (JPL) and California Institute of Technology (Caltech) interns C. Bell, E. Linden and E. Rapaport for assisting with wirebonding the SOFTS devices and measurement data acquisition, and Summer Undergraduate Research Fellowships (SURF) program for supporting these students. This research was partly supported by the NASA award NNH18ZDA001N-APRA, the UK-STFC Astronomy Observation and Theory Small Award (Project No. ST/Y002210/1), STFC Quantum for Science Grant (ST/Y004973/1), Leverhulme Trust Visiting Professorship of P. Day and the European Union (ERC, SPARTAN-FX, 101169947). Views and opinions expressed are however those of the author(s) only and do not necessarily reflect those of the European Uniopn or the European Research Council. Neither the European Union nor the granting authority can be held responsible for them. This work was performed in part at the JPL, which is operated by the California Institute of Technology under contract with the National Aeronautics and Space Administration (NASA). For the purpose of Open Access, the author has applied a CC BY public copyright licence to any Author Accepted Manuscript version arising from this submission.

\end{document}